\documentclass[sigplan,nonacm]{acmart}

\AtBeginDocument{}

\usepackage{booktabs}
\usepackage[normalem]{ulem}
\usepackage{tikz}
\usetikzlibrary{positioning,fit,arrows.meta}
\graphicspath{{figs/}}

\newcommand{\sysname}{\textsc{KREX}}
\newcommand{\dobench}{\texttt{do\_bench}}

\newcommand{\PHM}[1]{\vspace{.4em} \noindent\textbf{#1}\hspace{.5em}}

\usepackage{enumitem}
\definecolor{NatRed}{HTML}{C72228}
\newcommand{\leadpara}[1]{\noindent\textbf{#1}\hspace{.5em}}
\newcommand{\middlepara}[1]{\vspace{.3em}\noindent\textbf{#1}\hspace{.5em}} 
\newcommand{\stepnum}[1]{%
    \tikz[baseline=(char.base)]{%
        \node[shape=circle, fill=black, inner sep=1pt, text=white] (char) {\sffamily\bfseries\small#1};%
    }%
}

\begin{document}

\title[\sysname{}]{\sysname{}: Concurrent Kernel Benchmarking on Shared GPUs via
       Region-Granular Exclusivity}

\author{Tianyu Feng\textsuperscript{\textasteriskcentered}}
\affiliation{%
  \institution{HKUST}
  \country{Hong Kong SAR, China}
}

\author{Haoxuan Yu\textsuperscript{\textasteriskcentered}}
\affiliation{%
  \institution{HKUST}
  \country{Hong Kong SAR, China}
}

\author{Tianyuan Wu}
\affiliation{%
  \institution{HKUST}
  \country{Hong Kong SAR, China}
}

\author{Lingyun Yang}
\affiliation{%
  \institution{Alibaba Group}
  \country{China}
}

\author{Daocheng Ying}
\affiliation{%
  \institution{Alibaba Group}
  \country{China}
}

\author{Yuxiao Wang}
\affiliation{%
  \institution{Alibaba Group}
  \country{China}
}

\author{Ruibo Fan}
\affiliation{%
  \institution{Alibaba Group}
  \country{China}
}

\author{Yinghao Yu}
\affiliation{%
  \institution{Alibaba Group}
  \country{China}
}

\author{Guodong Yang}
\affiliation{%
  \institution{Alibaba Group}
  \country{China}
}

\author{Liping Zhang}
\affiliation{%
  \institution{Alibaba Group}
  \country{China}
}

\author{Wei Wang}
\affiliation{%
  \institution{HKUST}
  \country{Hong Kong SAR, China}
}

\renewcommand{\shortauthors}{Feng et al.}

\begin{abstract}

LLM agents automate GPU kernel optimization by repeatedly composing candidates
and measuring their duration on real GPUs. Existing systems
preserve measurement fidelity by reserving a GPU for an entire agent session or
benchmarking command. However, this results in poor utilization because only a
small fraction of command execution requires exclusive GPU access. Sharing GPUs
could recover this idle capacity, but introduces contention that compromises
measurement fidelity and misdirects the agent's search.

We present \sysname{}, a runtime for concurrent kernel agent benchmarking with
\emph{region-granular exclusivity}. \sysname{} lets agents mark \emph{critical
regions} involving timing-sensitive operations within a benchmarking command.
The runtime then enforces exclusivity within marked regions and allows concurrent execution outside them, achieving high throughput while preserving measurement fidelity.
To enforce \emph{in-region exclusivity}, \sysname{} blocks
new competing GPU submissions and drains outstanding work before freezing
sibling processes and isolating CPU cores, protecting both GPU execution and the
host threads that drive measurements. To maximize \emph{off-region concurrency},
\sysname{} reuses GPU contexts in persistent context processes to avoid
repeated, node-wide serialized context creation. We evaluate \sysname{} on
NVIDIA and AMD GPUs. Compared with command-granular exclusivity baselines,
\sysname{} delivers up to $3.4\times$ the benchmarking throughput with a negligible
p95 timing inflation of $0.30\%$, $1.58\%$, and $3.90\%$ for kernels longer than
10\,ms, 1\,ms, and 0.1\,ms, respectively.

\end{abstract}

\maketitle
\renewcommand{\thefootnote}{\fnsymbol{footnote}}%
\footnotetext[1]{Tianyu Feng and Haoxuan Yu contributed equally to this work.}%
\renewcommand{\thefootnote}{\arabic{footnote}}%

\section{Introduction}
\label{sec:intro}

Optimizing GPU kernels can significantly reduce the end-to-end computational
cost of modern AI
workloads~\cite{flashattention-neurips22,vllm-sosp23,datamovement-mlsys21}, but
it requires extensive expert knowledge and repeated cycles of profiling and
search, and the effort recurs with every hardware
generation~\cite{flashattention3-neurips24,halide-autoscheduler-siggraph19}.
Kernel agents automate this loop, generating implementations that rival, or even
surpass, hand-written
kernels~\cite{alphaevolve-arxiv25,cudaagent-arxiv26,cudal2-arxiv25}. Industry
now trains, benchmarks, and deploys kernel agents at
scale~\cite{fastkernels-arxiv26,sol-kernelagent-arxiv26}, as reported by
Meta~\cite{kernelevolve-arxiv25}, Google~\cite{alphaevolve-arxiv25}, and
public leaderboards~\cite{kernelbot-codeml25}. An agent improves a kernel by
measuring it, so training, validating, and using these agents all reduce to
running candidate kernels on real GPUs. In our production environment, users
submit over 300,000 of these \emph{kernel benchmarking} jobs daily across six
GPU models from three vendors, making kernel benchmarking a significant and
growing GPU workload.

To measure the performance of a candidate kernel, a benchmarking command is
launched to produce a timing result. Under the hood, it imports the frameworks
the kernel needs, compiles the kernel, generates inputs and checks the result
against a reference, and only then \emph{times} the kernel in a timing loop.
These phases make different demands on the device. Imports and compilation do
not touch the GPU, and input generation and correctness checking produce a value
rather than a timing. Only the timing loop produces a measured kernel duration, which
is the agent's optimization objective. To ensure the measurement is faithful,
kernel benchmarking systems run candidates on GPUs they reserve
\emph{exclusively}, without any interference from other workloads.

\begin{figure}[t!]
  \centering
  \includegraphics[width=\columnwidth]{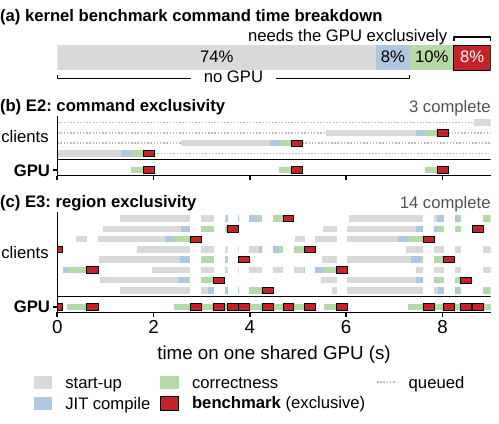}
\caption{{Granularity determines what exclusivity costs.} \textbf{(a)} Execution
time breakdown of a GEMM kernel benchmarking command. Only the timed loop needs
the GPU exclusively. \textbf{(b)} E2: whole candidate kernels hold the device.
\textbf{(c)} E3 (\sysname{}): only timed regions hold it exclusively.}
  \label{fig:granularity}
\end{figure}

However, exclusive device reservation results in low utilization. In our
production platform, queued kernel benchmarking requests alone demand
4.8$\times$ the online GPU capacity during peak hours, yet the GPUs that are
allocated remain idle for the majority of their allocation time. Existing
systems grant benchmarking exclusivity at two granularities, both far coarser
than the timing loop. 
\textbf{E1}: \emph{session-granular exclusivity} reserves
a GPU for the entire duration of an agent's session. We profiled a
representative kernel agent~\cite{atrex-arxiv26,atrex-agent-repo} over a complete session and
measured a GPU utilization rate of just 3.4\%. GPU commands occupy only 19.3\%
of the session's total wall-clock time, and even within those commands, GPU
activities are sparse, accounting for only 17.4\% of the execution time on average; the rest of
the session is LLM reasoning and other activities that use no GPU. 
\textbf{E2}:
\emph{command-granular exclusivity} reserves the GPU only while a command is
running, so multiple agent sessions share a GPU on demand. Public leaderboards
and internal deployments work this
way~\cite{kernelbot-codeml25,kernelevolve-arxiv25}. While it results in better
utilization than \textbf{E1}, the reservation remains much coarser than the work
it serves. In a run of eight agents contending for one GPU, the device was held
for 82\% of a nine-second window but spent only 7\% of that window executing the
timed regions (Figure~\ref{fig:granularity}b), because the timed loop accounts
for just 8.5\% of a command's total duration (Figure~\ref{fig:granularity}a).

One remedy for the low utilization is to let multiple commands share a GPU.
However, existing GPU-sharing mechanisms such as MPS, MIG, and
time-slicing~\cite{nvidia-mps,nvidia-mig} cannot be adopted directly. These
mechanisms are designed for throughput and isolation, but not for performance
fidelity, as they either perturb the timing loop's measurements or, in the case
of MIG, alter the hardware under test (\S\ref{sec:gpu-sharing-problem}). 

We argue that the conflict between utilization and fidelity is an artifact of
granularity. A command's reported kernel duration is sensitive to contention only within
small, identifiable \emph{regions}, which in practice correspond to its timing
loop; the rest of the command needs no exclusive access. This motivates a finer
level of exclusivity, \textbf{E3}: \emph{region-granular exclusivity}, which
makes the device exclusive inside a critical region and shared elsewhere
(Figure~\ref{fig:granularity}c), achieving high utilization while preserving
measurement fidelity. 

In this paper, we present \sysname{}, a runtime for concurrent
\underline{k}ernel benchmarking with \underline{r}egion-granular
\underline{ex}clusivity. \sysname{} exposes a simple client API that lets
the agent mark a critical region with a single line of
code, while leaving the kernel code and benchmarking scaffold unchanged.
It then uses two mechanisms to achieve measurement fidelity and high utilization:

\middlepara{In-region exclusivity.} Once the region is identified by an agent,
in-region exclusivity must be enforced despite other tenants' ongoing
work, including asynchronous GPU tasks already queued on the device. Not only
must this in-flight GPU work be excluded, but so must the tenants' host threads: for
multi-kernel implementations, the CPU operations feeding the GPU can themselves
sit on the critical path, making host-side contention an equally potent source
of interference. A critical region therefore has to own two resources at once, the
device and the host side that drives it. To achieve this, \sysname{} employs a
series of cooperative GPU- and CPU-side exclusivity mechanisms. Upon entering a
region, \sysname{} drains the in-flight GPU work with a protocol that leaves
nothing submitted on the device, freezes other tenants' host-side process trees,
and then executes the candidate kernel, along with its GPU context, on cores reserved
within that GPU's \emph{CPU slice}, a fixed partition of the host's cores. 

\middlepara{Off-region concurrency.} Work outside a critical region has no
exclusivity requirement and should run at the highest possible concurrency to
maximize throughput. In practice, however, that concurrency is limited by GPU
initialization and context creation, which acquire a driver-level lock that
blocks concurrent operations on \emph{all} GPUs in the node; in our testbed
these operations take hundreds of milliseconds for each command. 
To eliminate this recurring cost, \sysname{}
maintains a pool of pre-allocated, persistent contexts, each held by a
long-lived \emph{context process}, and reuses them instead of
creating one per command. \sysname{} interposes on the driver API and forwards a
command's calls to its assigned context process. These
calls are submitted asynchronously, overlapping call forwarding with GPU execution to hide the
added IPC overhead. Per-command cleanup and fault recovery maintain isolation
between commands.

We evaluate \sysname{} by replaying over 20,000 benchmarking commands
collected from three kernel agents running on three benchmark suites,
using a four-node testbed with 16 NVIDIA H20 and 16 AMD MI308X GPUs.
Compared with command-granular exclusivity, \sysname{} delivers
$3.4\times$ and $2.6\times$ the benchmarking throughput on H20 and MI308X,
respectively, while keeping p95 timing inflation at $0.30\%$, $1.58\%$,
and $3.90\%$ for kernels longer than 10, 1, and 0.1\,ms, respectively.
A candidate-ranking experiment on H20 shows that \sysname{} keeps
ranking-flip rates within three percentage points of an uncontended
repeat, preserving the relative performance signals that guide
the agent's search.

\section{Background and Motivation}
\label{sec:background}

\subsection{Kernel Agent and Benchmarking}
\label{sec:motivation-loop}

\leadpara{Kernel agent.}
Kernel agents generate candidate kernels and evaluate them based on measured
kernel duration, which guides their search for better-performing implementations. These
kernels are often benchmarked in Python environments, such as PyTorch, where the
reference implementation and input tensors already
reside~\cite{kernelbench,kernelbot-codeml25}. A typical benchmarking command
imports the required frameworks, compiles the candidate kernel, generates inputs
and checks correctness against the reference, and finally times the kernel.
Timing follows standard practices: benchmarking harnesses execute warm-up
launches, followed by repeated timed launches, and summarize the results with a
statistic such as the median, often flushing the L2 cache between iterations.

\middlepara{Benchmarking command.}
The phases of a benchmarking command differ in whether they need the device at
all, and in whether they need it to themselves. Framework imports and kernel
compilation do not involve the GPU. Input generation and correctness checks
utilize the GPU but do not require exclusive access: colocated workloads may
delay their completions, but do not change the tensors they generate or the
verdict they return. Only the timing phase requires \emph{exclusive hardware
access}, as its output directly determines the kernel-duration metric the agent
optimizes. 

Compared to other phases, the timing phase is often a small fraction of the
command that contains it. In a representative kernel agent workload on our
testbed, start-up and compilation account for 81\% of a command's median
duration, the correctness check for 10\%, and the timed loop for only 8.5\%, or
0.22\,s out of a 2.6\,s command (Figure~\ref{fig:granularity}a). 

\subsection{Why Current Approaches Fail}
\label{sec:current-approaches-fail}

\leadpara{Reservation harms utilization.}
Existing systems protect measurement fidelity by reserving each GPU at one of
two granularities. \textbf{E1}: \emph{session-granular exclusivity} holds a GPU
for an agent's entire session. In one profiled session of a representative
kernel agent~\cite{atrex-agent-repo}, GPU commands occupy 19.3\% of wall-clock
time, and the device is active for only 17.4\% of that interval, yielding 3.4\%
utilization. LLM reasoning and other non-GPU work fill the remainder.
\textbf{E2}: \emph{command-granular exclusivity} holds the GPU only while each
benchmarking command runs, as public leaderboards and internal deployments
do~\cite{kernelbot-codeml25,kernelevolve-arxiv25}. This removes the session-wide
reservation but still holds the device for the entire command, although only the
timing phase requires exclusivity, which accounts for a small fraction of the
command time (\S\ref{sec:motivation-loop}). 

\begin{figure}[t]
  \centering
  \includegraphics{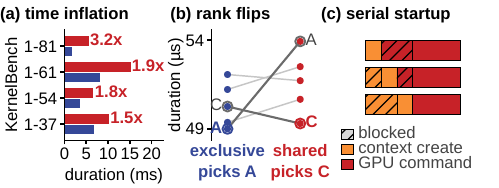}
\caption{
\textbf{(a)} GPU sharing inflates measurements. \textbf{(b)} Contention distorts ranks.
\textbf{(c)} Node-wide serialization caused by context creation. 
}
  \label{fig:motivation}
\end{figure}

\label{sec:motivation-waste}
Command-granular exclusivity also results in a recurring GPU setup cost upon
command start. Each command runs in a fresh process, initializes the GPU and
creates its own context inside the reservation. These operations acquire a
driver-level lock that blocks operations on every GPU in the node, so only one
command can perform them at a time; each costs hundreds of milliseconds. In our
experiments, creating 64 CUDA contexts on one server takes more than 30 seconds
even when the requests are issued concurrently across different GPUs. Command
admission therefore serializes node-wide; as commands shorten, waiting to start
consumes a larger share of each reservation (Figure~\ref{fig:motivation}c).

\label{sec:gpu-sharing-problem}
\middlepara{Sharing compromises fidelity.}
One way to improve GPU utilization is to time-multiplex
commands~\cite{nvidia-mps}, but unprotected sharing compromises measurement
fidelity. Concurrent agents interleave kernels on the same GPU, so contention
makes measured kernel duration depend on colocated workloads rather than the kernel
alone. Figure~\ref{fig:motivation}a shows the effect: replaying the same benchmark
commands with eight per GPU and no protection inflates their reported kernel durations
to $1.5\times$--$3.2\times$ those measured alone.
As such load-dependent distortion introduced by contention tends to inflate measured
kernel duration, repeating measurements under contention does not by itself recover
uncontended kernel duration. Moreover, since the distortion varies with colocated workloads,
it cannot be reliably canceled in candidate comparisons.

Kernel agents are especially vulnerable because they rank candidates by relative
performance. In one trajectory, an agent optimizing an MoE token-alignment
kernel produced several candidates within a few percent of one another
(Figure~\ref{fig:motivation}b). Without contention, its best candidate took
49.0\,$\mu$s, 2.6\% less time than its closest rival at 50.3\,$\mu$s. Under
unprotected sharing with 64 agents, their ranking reversed: the best candidate
measured 53.9\,$\mu$s versus the rival's 49.3\,$\mu$s, so an agent reading those
numbers keeps the slower kernel. As shown in Figure~\ref{fig:motivation}b,
sharing reversed six of the ten pairwise rankings among these candidates,
whereas a second uncontended run reproduced the original ordering.

\subsection{Region-Granular Exclusivity}
\label{sec:motivation-idea}

\begin{figure*}[t]
  \centering
  \includegraphics{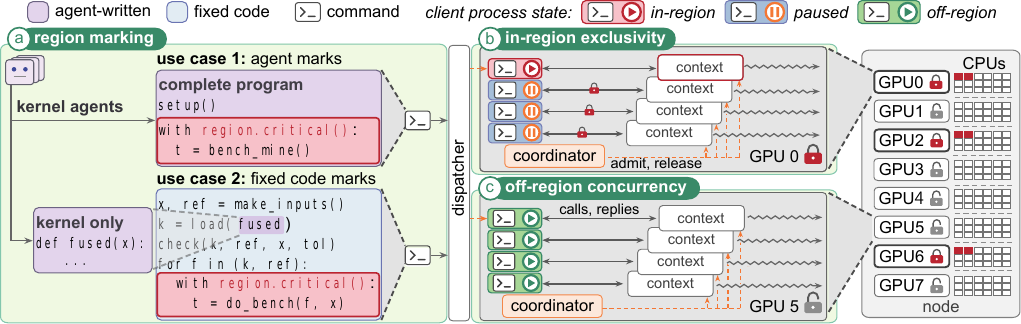}
\caption{Overview of \sysname{}, shown with eight GPUs and four concurrent commands per GPU for illustration.
}
  \label{fig:arch}
\end{figure*}

\leadpara{Region-granular exclusivity.}
Existing approaches make exclusivity all-or-nothing at command granularity:
reserve the whole command or share all of it. Yet only the timing phase of a
command is sensitive to contention (\S\ref{sec:motivation-loop}). Confining
exclusivity to the timing phase is therefore \emph{sufficient} for fidelity,
since the reported kernel duration depends on nothing that happens outside it. Doing so
is also \emph{cheap}, as the timing phase usually occupies only a small fraction of
a command (\S\ref{sec:motivation-loop}). Formally, we call the marked span of a
command that must run alone a \emph{critical region}, and enforcing
exclusivity over that span \emph{region-granular exclusivity} (\textbf{E3}).
\textbf{E3} preserves the protected measurement interval of \textbf{E1} and
\textbf{E2} while leaving the rest of the command available
to concurrent execution (off-region concurrency).

\middlepara{Efficiency challenge.}
To fully exploit off-region concurrency, the system must ensure that enough
kernel candidates become GPU resident quickly. Under the existing per-command process
model, however, each command first initializes the GPU and creates a context. These
operations acquire a driver-level lock shared across every GPU in the node, so
commands initialize one at a time even when they target different GPUs: while
one command creates its context, the others wait on assigned but idle GPUs
(Figure~\ref{fig:motivation}c). Command admission therefore remains serialized,
significantly limiting off-region concurrency.

\middlepara{Requirements.}
Building an efficient system that implements region-granular exclusivity for
concurrent and faithful kernel benchmarking has four requirements:

\begin{itemize}[leftmargin=1.2em, itemsep=1pt, topsep=2pt, parsep=0pt]
  \item \textbf{R1 (boundary).} The system should provide a simple interface 
  for the agent or scaffold author to mark the complete region
  without modifying the kernel under test, and enforce the supplied boundary
  throughout the command.
  \item \textbf{R2 (fidelity).} The system should keep reported timings close to
  uncontended runs and limit distortion of candidate rankings.
  \item \textbf{R3 (throughput).} The service must keep multiple commands
  resident per GPU and overlap their off-region work without allowing node-wide
  GPU initialization and context creation to serialize admission.
  \item \textbf{R4 (heterogeneity).} The same region interface and isolation
  contract must hold on diverse hardware, specifically NVIDIA and AMD GPUs, even when
  their enforcement mechanisms differ.
\end{itemize}

\section{System Overview}
\label{sec:overview}

\sysname{} is a multi-tenant runtime for concurrent kernel agent benchmarking
that provides region-granular exclusivity. It accepts commands whose programs
contain explicitly marked critical regions. During these regions, the runtime
protects GPU execution and CPU cores from competing tenants. When no command on
a GPU is executing a marked region, the runtime allows commands on that GPU to
execute concurrently. Users only mark the code that needs exclusivity while the
runtime manages resource allocation and coordinates exclusive access.

\middlepara{Workflow.}
Figure~\ref{fig:arch} follows a command from region marking to execution.
Regions are marked before command submission in one of two ways
(Figure~\ref{fig:arch}a). In the \emph{agent-marked} case, the agent writes the
complete program, including the benchmarking script, and marks its regions. In
the \emph{scaffold-marked} case, the agent supplies only the kernel; the
scaffold author marks the regions in a fixed benchmarking script that the agent
cannot modify.

Each command runs in a \emph{client process}, and the \emph{dispatcher} pairs it with
an available \emph{context process} from the pool. The client executes the
candidate program and forwards its GPU calls to the context process, which holds
the GPU context and submits those calls to the driver. The context process
persists across commands, allowing its GPU context to be \emph{reused} rather than
recreated for each command.

When a command enters a region, \sysname{} first blocks new GPU submissions from
other tenants on that GPU, then drains their outstanding GPU work, and finally
freezes their process trees (Figure~\ref{fig:arch}b). Physical CPU cores are
partitioned into \emph{disjoint per-GPU slices} throughout execution, preventing
tenants on different GPUs from competing for the same cores. Within a slice, the
measuring command's submitting threads are pinned to reserved cores before
timing. On region exit, the runtime thaws the other tenants and reopens
submissions, allowing them to resume execution. Concurrent execution is allowed
when no command on a GPU is in a region (Figure~\ref{fig:arch}c). Context reuse
sustains this concurrency by avoiding repeated, serialized GPU context creation
(\S\ref{sec:motivation-waste}).

These mechanisms address the requirements in \S\ref{sec:motivation-idea} as
follows. \textbf{R1 (boundary):} \emph{Region marking} lets agents and scaffold
authors express measurement boundaries through a one-line API
(\S\ref{sec:design-regions}). \textbf{R2 (fidelity):} \emph{In-region
exclusivity} excludes competing GPU work and protects the CPU threads that
submit the measurement's GPU calls (\S\ref{sec:design-entry}). \textbf{R3
(throughput):} \emph{Off-region concurrency} overlaps commands outside regions,
while the persistent context pool pays context creation costs at startup or
recovery rather than on every command (\S\ref{sec:design-pool}). \textbf{R4
(heterogeneity):} Vendor-specific adaptations implement the shared region
protocol on NVIDIA and AMD (\S\ref{sec:impl}).
\section{Design Mechanisms}
\label{sec:design}

In this section, we describe the design mechanisms in detail, following
the roadmap given in \S\ref{sec:overview}.

\subsection{Region Marking}
\label{sec:design-regions}

In \sysname{}, critical regions must be marked before command submission. This
is done with a simple context manager, \texttt{region.critical()}. \sysname{}
then establishes exclusivity for the enclosed code at runtime
(\S\ref{sec:design-entry}).

\middlepara{Why boundaries cannot be inferred.}
A region should hold all measurement code and nothing more: too broad, it
reserves the GPU for work that tolerates sharing and costs concurrency; too
narrow, it leaves measurement code exposed to contention and reports a distorted
duration. Where that boundary falls differs across commands. Marking only the
code between timing calls excludes the warm-up loop, which precedes the timed
loop to complete hidden initializations and stabilize cache state; marking every
loop to capture the warm-up pulls in correctness checks. Region sizes span
orders of magnitude, from a loop that relaunches a single kernel to an entire MoE
forward pass of many kernels. Some measurements are implicit in the calling code,
as when an autotuner profiles internally, and never appear as a timed loop. The
boundary follows the intended measurement, which the code does not state.

\middlepara{Why authors supply them.}
We therefore require explicit region marking given by the code author, rather
than infer regions automatically from code that may not capture the author's
measurement intent. Region marks are supplied in two ways
(Figure~\ref{fig:arch}a). When an agent writes the complete program, including
both the kernel and the benchmarking script, it also marks the regions. When a
fixed benchmarking scaffold is used to prevent the agent from modifying the
evaluation procedure, the agent submits only the generated kernel, and the
scaffold author marks the regions in advance. Letting code declare what a lower
layer cannot infer is established practice: Linux's \texttt{madvise}~\cite{man-madvise}
and \texttt{posix\_fadvise}~\cite{man-posix-fadvise} pass down access patterns,
and applications have long disclosed I/O intent to guide prefetching and caching
decisions a file system cannot make on its
own~\cite{informed-prefetching-sosp95,app-controlled-caching-osdi94}.

Letting agents mark regions also keeps a wrong boundary where it can be
repaired. As the boundary reflects the agent's intended measurement, if the
results contradict what the agent expected, it may revise the mark, just as it
repairs a kernel it wrote. By contrast, a boundary derived by an external tool
is invisible to the agent, which charges the resulting measurement error to its
composed kernel and has no mark to correct. In the absence of a reliable
analytical method for identifying region boundaries, \sysname{} therefore relies
on the author's marks in this setting.

\subsection{In-Region Exclusivity}
\label{sec:design-entry}

\begin{figure}[t]
  \centering
  \includegraphics{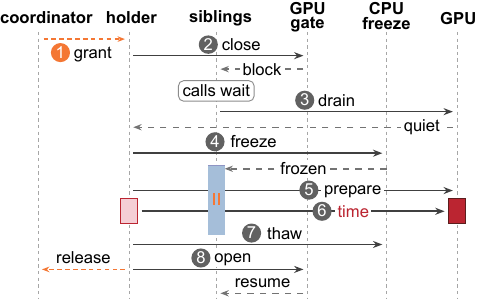}
\caption{Region entry and exit on one GPU.}
  \label{fig:design-entry}
\end{figure}

When a tenant enters a region, other tenants may still have GPU work executing
or queued, and may continue submitting more. The region must exclude both
previously submitted work and new submissions. Moreover, device exclusivity
alone is insufficient as \emph{contention on the CPU} can delay the threads that submit
successive GPU operations and kernels, increasing the measured kernel duration as well.
Therefore, \sysname{} combines GPU exclusivity with CPU protection for the
threads that drive the measurement. Figure~\ref{fig:design-entry} walks through
this protocol in eight steps, which we reference by circled number for the
remainder of this subsection. For simplicity, we refer to the tenant granted
region admission as the \emph{holder}, while other tenants on the same GPU are
its \emph{siblings}. The client process runs the candidate program submitted for
benchmarking; on NVIDIA, a driver shim forwards its GPU calls to a context
process that owns the GPU context (\S\ref{sec:design-pool}). We describe the GPU
protocol in terms of these context processes; on AMD, a runtime-loaded hook
executes the same protocol directly in the client process (\S\ref{sec:impl}).

\middlepara{GPU exclusivity.} A region requires that no other tenant on the same
GPU submit work while it runs and that all work those tenants submitted earlier
has completed before it starts. To enter a region, the tenant's context process
acquires a region lock from its GPU's \emph{coordinator}, which admits one
tenant at a time~\stepnum{1}. The holder's context process then atomically
claims a shared-memory \emph{gate}, recording the holder's identity and a unique
\emph{epoch} for this entry~\stepnum{2}. Each context process registers before
initializing or submitting GPU work so the holder can identify all processes to
drain. Registration is serialized with gate closure: processes admitted before
closure are included in the drain, while later arrivals must wait for the gate
to reopen before using the GPU. The registration entries for the processes being
drained cannot be reassigned to replacement processes until the region ends.

To facilitate draining, each context process maintains an \emph{in-flight
counter} for GPU operations in the host submission path. Before issuing an
operation that may leave work on the GPU (e.g., a kernel launch or async
\texttt{memcpy}), the submitting thread increments the counter and checks the gate. If
another tenant is holding the gate, it retracts the increment and parks until
the gate state changes, issuing nothing and consuming no CPU while blocked.
Otherwise, it issues the call and decrements the counter on return, including
error returns. With ordered atomic accesses, incrementing before checking the
gate ensures that a racing submission is either covered by the drain or blocked
before reaching the driver. Context creation and cleanup follow the same
\emph{counter-and-gate} protocol: the counter makes the drainer wait for admitted
context changes to finish, while the gate prevents siblings from creating or
cleaning up contexts during the region.

After closing the gate, the holder's context process wakes a background drainer thread in each sibling context process, independently of client activity~\stepnum{3}.
Each drainer captures the epoch and waits for its process's counter to reach zero.
A zero counter only indicates that host submission has finished while asynchronously submitted GPU work may still be queued or executing.
The drainer therefore synchronizes all contexts owned by its process, including candidate-created secondary contexts, keeping their handles valid until synchronization returns.
After all context synchronizations succeed, the drainer acknowledges completion by publishing the captured epoch in shared memory.
The holder's context process polls these acknowledgements and the in-flight counters, proceeding only when every sibling has acknowledged the current epoch and has a zero counter.
Matching epochs prevents a previous region's confirmation from being mistaken for completion of newly submitted work.

\middlepara{CPU exclusivity.} Blocking GPU submissions alone does not stop
siblings from importing frameworks, compiling kernels, or generating inputs on
the CPU. However, these activities can interfere with in-region measurement by
preempting the holder's GPU-submitting threads, delaying GPU submissions.
\sysname{} therefore freezes sibling process trees once their GPU work has
drained~\stepnum{4}. The drain must finish before siblings are frozen so their
threads can complete outstanding driver calls and context synchronization.
Specifically, each candidate is placed in a per-tenant freezer \texttt{cgroup}
at process creation, allowing \sysname{} to pause the client and its
subprocesses without waiting for client-side registration. For example, a client
may execute \texttt{import} \texttt{torch} before registering with the
coordinator. The \texttt{cgroup} allows \sysname{} to pause the client during
this import, whereas a scheme that freezes only registered clients would leave
it running.

Tenants assigned to other GPUs must also be kept from competing for the holder's
cores without freezing their process trees. Therefore, in \sysname{}, each GPU
is assigned a \emph{disjoint slice} of physical CPU cores local to its NUMA node, and
all of its tenants are confined to that slice. Within each slice, dedicated
cores are reserved for the threads that submit GPU work when any tenant is
in-region. This fixed partition prevents a tenant with high CPU demands from
running on idle cores assigned to another GPU and interfering with its in-region
measurement. The slice size determines the CPU capacity available to tenants of
each GPU and is configured at deployment. \S\ref{sec:impl} describes how these
CPU boundaries are enforced.

\middlepara{Region entry and exit.} After completing the GPU and CPU isolation
steps above, \sysname{} performs the final preparation for
measurement~\stepnum{5}. It flushes L2 cache with a \texttt{memset} over an L2-sized
buffer and pins the holder's submitting threads to the cores reserved within its
GPU's CPU slice. The flush removes cache state left by siblings, while pinning
protects the submission threads from CPU contention that could inflate the
measured kernel duration. If synchronization fails, the entry request is canceled or
times out, gate ownership is lost, or a participant dies during entry, the
request fails without starting measurement. Otherwise, the holder proceeds to
execute the timed code~\stepnum{6}.

Normal region exit begins when the timed code finishes.
The holder's context process first waits for all GPU calls during the region to finish host-side submission.
After that, it thaws its siblings~\stepnum{7}.
Only after thawing is complete does it reopen the gate and release the region lock~\stepnum{8}.
If the lock were released before thawing finished, the next holder could acquire it and freeze the same sibling processes.
The previous holder's delayed thaw could then undo the new holder's freeze, allowing those processes to compete for CPU time during the new holder's measurement.

If the holder dies before completing this exit sequence, the \emph{dispatcher}
(Figure~\ref{fig:arch}) takes over cleanup. It first confirms that the holder's
admitted GPU work has ended, then thaws the siblings frozen by that holder
before reclaiming the gate. To prevent recovery from racing with normal exit,
\sysname{} serializes the two paths for each context process so they cannot
modify the gate and freeze state concurrently. It also binds cleanup to the
original holder and entry epoch, preventing delayed recovery from reopening a
later entry's gate or undoing its CPU freeze. After a failed entry or a holder's
death, reclaiming the gate alone does not make the GPU ready for a new
measurement. \sysname{} must first finish cleanup and restore the affected GPU
resources to a usable state. If a context becomes unusable, \sysname{} retires
and replaces it as part of recovery (\S\ref{sec:design-pool}).

\subsection{Off-Region Concurrency}
\label{sec:design-pool}
\label{sec:design-arch}

Region-granular exclusivity allows the work outside one candidate's regions to
concurrently execute with work from other candidates on the same GPU. However, a
command that creates its own GPU context pays for GPU initialization and context
creation, which acquire a driver-level lock shared across the node. These
operations take hundreds of milliseconds per command, and the shared lock
serializes them even across different GPUs (\S\ref{sec:motivation-waste}).
Simply running more commands concurrently therefore does not remove this
bottleneck if each command creates its own context.

\middlepara{Context pool.} To eliminate this recurring initialization tax,
\sysname{} reuses GPU contexts across commands by separating their lifetime from
that of candidate programs. Each candidate still runs in its own client process.
Each GPU context is held by a dedicated \emph{context process} that remains
alive across commands and serves candidates sequentially. \sysname{} creates a
fixed pool of these contexts and their owning processes for each GPU at startup
instead of creating a new context for every command. The client process can exit
independently of the persistent context process, allowing the context to be
reused after the candidate exits.

For each incoming command, the dispatcher pairs its client process with a free context process on the least-loaded GPU.
Pairing occurs only at candidate boundaries and requires a control-plane round trip rather than context creation.
Once the pair is connected, GPU calls do not pass through the dispatcher.
If the pool is full, the dispatcher parks the client server-side rather than requiring repeated acquisition requests.
This avoids the excessive acquisition requests that halved region throughput in an early deployment.

A shim in the client process interposes on the driver API, including dynamically
resolved driver entry points, and forwards calls to the paired context process
through shared-memory remote procedure calls (RPCs). This interposition
preserves compatibility with unmodified candidate code. Unsupported calls fail
explicitly rather than bypassing the pool. The context process also enforces GPU
exclusivity for that candidate, as described in \S\ref{sec:design-entry}. This
separation removes repeated context creation from the normal command path.
However, context reuse introduces two challenges: RPC overhead can perturb
measurements, and retained device state requires cleanup and fault recovery
between commands.

\begin{figure}[t]
  \centering
  \includegraphics{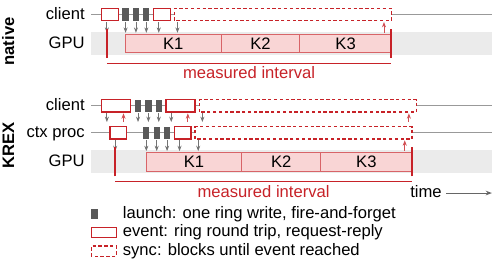}
\caption{One timed loop under native execution (top) and \sysname{} with asynchronous forwarding (bottom).}
  \label{fig:design-pool-fig}
\end{figure}

\middlepara{Preserving measurement fidelity.} RPC overhead can delay GPU
submissions and introduce gaps between kernels inside the timed loop. \sysname{}
reduces this overhead through three mechanisms:

\uline{\emph{1) Minimal shared-memory transport.}}
The client and context process communicate through two single-producer,
single-consumer shared-memory ring buffers. A kernel launch requires one
fixed-layout ring entry, while bulk data moves through registered shared-memory
areas accessible to both processes.

\uline{\emph{2) Asynchronous forwarding.}} \sysname{} reduces RPC overhead by
using one-way forwarding for calls that do not require a synchronous reply. For
example, kernel launches use fire-and-forget requests: the client submits the
request without waiting for a reply, allowing forwarding to overlap with GPU
execution (Figure~\ref{fig:design-pool-fig}). Timing events and synchronization
operations instead use request--reply communication. Asynchronous forwarding may
still add transport delay to the measured interval, but this delay is typically
negligible relative to the interval.

\uline{\emph{3) Spin--park waiting policy.}} Under this policy, threads spin
briefly for a ring entry and park on a \texttt{futex} only if no entry arrives within the
spinning interval. An entry that arrives during the spinning interval can be
consumed immediately without a thread wake-up that costs several microseconds.
Blocking on every wait would add percent-level measurement overhead to a
launch-dense timed iteration due to frequent wake-ups. If the wait persists over
a given time threshold, the thread parks to yield CPU resources.

\middlepara{Context cleanup and fault recovery.} Context reuse carries device
state across command boundaries, and the context process must explicitly clean
up that state upon command exit. After a candidate finishes, its
context process releases all of the candidate's allocated resources such as
memory allocations, host registrations and loaded modules. \sysname{} retains
the GPU context and a content-addressed library cache, and then synchronizes the
device to check whether the context remains usable. If this health check fails
because of an unhandled error such as a sticky launch error or an unfinished
graph capture, the dispatcher retires and replaces the context process rather
than assigning it another candidate. A background check also detects clients
that terminate unexpectedly or exceed their execution time limit, in which case
it applies the same recovery rule to the context processes they paired with. If
a candidate completes normally and cleanup finishes without errors, the context
process will be reused to serve the next candidate. Therefore, a candidate
failure is confined only to that run and can be recovered by a simple
context-process respawn.

\section{Implementation}
\label{sec:impl}

\sysname{} supports NVIDIA and AMD GPUs with about 20{,}000 lines of C++ for
GPU-call interception and dispatch. The dispatcher, per-GPU coordinators, and
kernel-evaluation service use 4{,}800 lines of Python outside the GPU-call path.
Another 41{,}000 lines of C++ CUDA and NVML forwarding stubs are machine
generated from annotated headers to avoid hand-written marshalling for new entry
points.

The NVIDIA and AMD implementations follow the same region protocol in
\S\ref{sec:design-entry} but differ in interception, draining, and freezing. On
NVIDIA, the gate is implemented in the context process's RPC dispatch loop. On
AMD, the gate can be implemented either in the same pooled path or in a tools
library loaded by ROCm at startup inside the tenant process. The library patches
the runtime dispatch table to intercept each queue's kernel launches. \sysname{}
provides both implementations and exposes them as configuration options. Unlike
NVIDIA's host-submission counter, AMD's counter tracks device completion: the
counter is incremented on entry to each operation and decremented by a polling
thread when the operation's completion signal fires. The holder waits for
sibling counters to reach zero without sibling-side context synchronization or
dedicated drainer threads. On AMD, tenants are added to freezer cgroups only
after initialization because freezing during initial memory registration causes
failures on the evaluated system.
\section{Evaluation}
\label{sec:eval}

Our evaluation answers five questions:
\textbf{(Q1)} how long benchmarking commands run and what fraction of their execution time
is spent in critical regions
(\S\ref{sec:eval-workload});
\textbf{(Q2)} how much throughput \sysname{} gains over command-granular exclusivity
(\S\ref{sec:eval-throughput});
\textbf{(Q3)} how closely its timings and candidate rankings match uncontended execution
(\S\ref{sec:eval-fidelity});
\textbf{(Q4)} how individual components and configuration choices affect the
throughput--fidelity trade-off
(\S\ref{sec:eval-ablation}); and
\textbf{(Q5)} how GPU-call forwarding affects measured kernel duration
(\S\ref{sec:eval-limits}).

\subsection{Experimental Setup}
\label{sec:eval-setup}

\leadpara{Testbed.} We evaluate \sysname{} on two testbed platforms each
containing two 8-GPU nodes, for a total of 16 NVIDIA H20 and 16 AMD MI308X GPUs.

\middlepara{Workloads.} We evaluate \sysname{} by replaying more than 20k
    benchmarking commands (16k on NVIDIA and 4k on AMD) collected from real
    kernel-agent runs. To collect these traces, we integrate \sysname{} with
    Atrex Kernel Agent~\cite{atrex-arxiv26,atrex-agent-repo} and
    kernel-design-agents~\cite{kda-repo} on NVIDIA, and GEAK~\cite{geak-repo} on
    AMD. We run these agents on tasks from three kernel benchmarks:
    KernelBench (KB)~\cite{kernelbench},
    Atrex-Bench (AB)~\cite{atrex-arxiv26,atrexbench-repo}, and
    FlashInfer-Trace (FT)~\cite{flashinfer-trace}, and record every benchmarking
    command together with its associated source code for replay.
  
\middlepara{Baselines.} We compare four baselines. The \emph{native} baseline
    provides command-granular exclusivity by running one command per GPU without
    \sysname{}. \emph{\sysname{}-serial} runs one command per GPU through
    \sysname{} to assess the runtime's effect without concurrent sharing. The
    full \emph{\sysname{}} configuration runs concurrent commands with
    region-granular exclusivity. The \emph{ungated} configuration runs
    concurrent commands without region protection to evaluate throughput and
    measurement fidelity under unprotected sharing. We define \emph{command
    concurrency} as the number of concurrent commands per GPU. Unless otherwise
    specified, \sysname{} and \emph{ungated} use a command concurrency of 16 on
    H20 and 4 on MI308X, with MPS~\cite{nvidia-mps} enabled for \sysname{} on H20.

\subsection{Workload Characterization}
\label{sec:eval-workload}

\begin{figure}[t]
  \centering
  \includegraphics{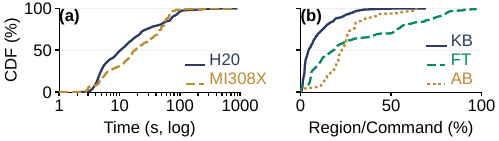}
\caption{Distributions of workloads. \textbf{(a)} Command durations on two GPU models (log x-axis);
\textbf{(b)} critical-region percentages of commands sampled on H20.}
  \label{fig:share}
\end{figure}

Figure~\ref{fig:share}a shows the distribution of command durations in the
recorded agent sessions, with medians of 10.2\,s on NVIDIA and 17.7\,s on AMD.
These durations reflect how long each command reserves the GPU under
command-granular exclusivity. To quantify the fraction of execution time spent
in critical regions, we also analyze 1069 commands from one hour of
\sysname{}-serial execution on H20. For each command, we sum the time spent in
all of its critical regions. Figure~\ref{fig:share}b shows that the median
region fraction is only $12\%$, and $73\%$ of commands spend less than $25\%$ of
their execution time in regions. Median fractions vary across suites but remain
below $25\%$: $4.4\%$ for KernelBench, $18.2\%$ for FlashInfer-Trace, and
$22.7\%$ for Atrex-Bench. These results indicate substantial off-region work
that \sysname{} can overlap across commands.

\subsection{Throughput}
\label{sec:eval-throughput}

\begin{figure}[t]
  \centering
  \includegraphics{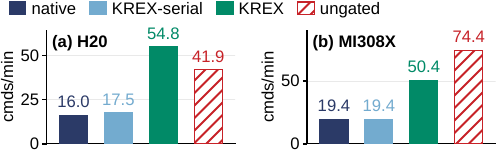}
\caption{Benchmarking throughput.}
  \label{fig:tput}
\end{figure}

To measure benchmarking capacity, we conduct one-hour saturated replays with
all baselines receiving the same command stream within each platform.
This replay excludes variability from agent reasoning and LLM response latency, allowing us to measure the benchmarking system's throughput in isolation. We report throughput per 8-GPU node.

On H20 (Figure~\ref{fig:tput}a), \sysname{} delivers 54.8 commands/min (cmds/min). This is
$3.4\times$ the throughput of native at 16.0 cmds/min and $3.1\times$ that
of \sysname{}-serial at 17.5 cmds/min. Note that \sysname{}-serial
outperforms native by about $9\%$ because context reuse removes repeated
context-creation overhead. \sysname{} additionally overlaps commands outside
their marked regions, where context reuse becomes more important at higher
concurrency, as creating more GPU contexts exacerbates the context-creation
bottleneck. Unexpectedly, ungated GPU sharing achieves
only 41.9 cmds/min at the same command concurrency, below \sysname{}'s
throughput despite running without protection. This is because \sysname{} uses
MPS to further improve off-region concurrency. We discuss the effect of MPS in
detail in \S\ref{sec:eval-ablation}.

On MI308X (Figure~\ref{fig:tput}b), \sysname{} delivers 50.4 cmds/min. This
is $2.6\times$ the throughput of native, which settles at 19.4 cmds/min.
\sysname{}-serial achieves nearly the same throughput as native because we
disable context pooling on MI308X, forgoing its throughput benefits to avoid the
greater fidelity loss it causes compared with H20. Ungated achieves a higher
throughput of 74.4 cmds/min ($3.8\times$ native), but at the cost of
substantially worse measurement fidelity, rendering its measurements unreliable
for kernel optimization.

\subsection{Measurement Fidelity}
\label{sec:eval-fidelity}

\begin{figure}[t]
  \centering
  \includegraphics{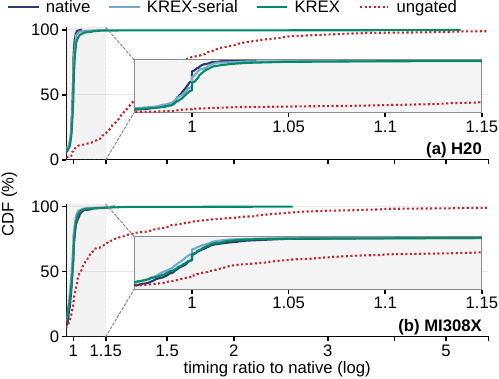}
\caption{CDFs of timing ratios to native for commands with native kernel
duration $>1$\,ms.
Main plots use log x-axes; insets show x=0.97--1.15 on linear axes.}
  \label{fig:cdf}
\end{figure}

\begin{figure}[t]
  \centering
  \includegraphics{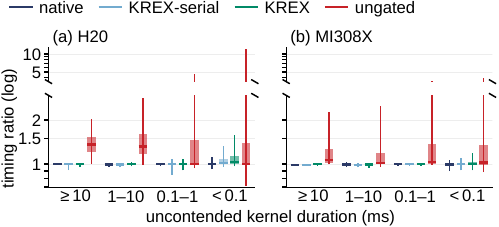}
\caption{Timing ratios by kernel duration (log y-axes);
boxes: p25--p75; whiskers: p05--p95; ticks: medians.}
  \label{fig:lenfid}
\end{figure}

High throughput is useful only if measurements remain faithful for kernel optimization.
We evaluate measurement fidelity by comparing reported kernel durations
and candidate rankings against native execution.
For kernel durations, we measure how much reported timings exceed their native
references using the p95 timing inflation, denoted by $P$, where
\begin{equation}
  P = \mathrm{p95}(r) - 1, \qquad
  r = \frac{t_{\mathrm{measured}}}{t_{\mathrm{native}}}.
  \label{eq:timing-inflation}
\end{equation}
Here, $t_{\mathrm{measured}}$ and $t_{\mathrm{native}}$ are each command's reported
kernel duration and native reference duration, respectively.
We report $P$ as a percentage; values closer to zero indicate less timing inflation
at the 95th percentile, capturing the upper tail rather than only the typical shift.
Because native measurements also vary across runs, we use a second native run
as a control to distinguish sharing-induced distortion from run-to-run variation.
We restrict the comparison in Figure~\ref{fig:cdf} to commands with native kernel
duration $>1$\,ms to reduce the influence of measurement noise, 
since even native execution
exhibits substantial run-to-run variation for shorter kernels.
We separately analyze measurement fidelity across kernel durations
in Figure~\ref{fig:lenfid}.

With the throughput configurations of Figure~\ref{fig:tput}, \sysname{} achieves
$P=2.5\%$ on H20 and $2.4\%$ on MI308X, compared with $154\%$ and $150\%$
for ungated, respectively (Figure~\ref{fig:cdf}).
For comparison, native and \sysname{}-serial have $P=0.9\%$ and $1.2\%$ on H20,
and $3.2\%$ and $1.6\%$ on MI308X, respectively.
Thus, \sysname{} combines the throughput gains in the preceding section with
p95 timing inflation close to that of native execution, whereas ungated sharing
substantially inflates measured durations.

\sysname{} preserves typical timings as well as limiting upper-tail inflation.
Median timing ratios for native, \sysname{}-serial, and \sysname{} span
0.997--1.000 on H20 and 0.993--0.997 on MI308X, compared with 1.331 and 1.032
for ungated, respectively.
Across the same three configurations, $93.6$--$97.1\%$ of timing ratios on H20
and $93.3$--$94.5\%$ on MI308X fall within $5\%$ of the native reference,
compared with $11.0\%$ and $53.8\%$ for ungated, respectively.

\middlepara{Fidelity across kernel durations.} Figure~\ref{fig:lenfid} groups timing
ratios by kernel duration under native execution on H20 and MI308X.
For the two duration bands at 1\,ms and above, \sysname{}'s $P$ ranges from
$1.2$--$3.0\%$ on H20 and $0.8$--$2.5\%$ on MI308X, while ungated exceeds
$100\%$ in every band on both platforms.
Shorter kernels require separate consideration: below 0.1\,ms, even the native
repeat has $P=12.1\%$ on H20 and $7.6\%$ on MI308X, so distortion at this scale
cannot be attributed solely to sharing.
\S\ref{sec:eval-limits} separately examines how RPC forwarding affects measured duration.

\begin{figure}[t]
  \centering
  \includegraphics{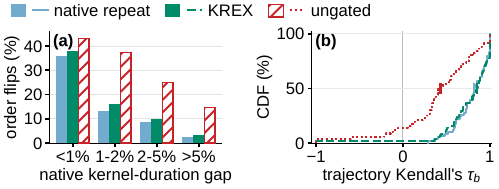}
\caption{\textbf{(a)} Pairwise ordering-flip rates grouped by relative
kernel-duration gap in the first native replay.
\textbf{(b)} CDFs of per-trajectory Kendall's $\tau_b$; ticks: medians.
Both panels use the first native replay as the reference.}
  \label{fig:rank}
\end{figure}

\PHM{Candidate rankings.}\label{sec:eval-rank}
Beyond individual timing accuracy, we assess whether measurements preserve
candidate rankings within an agent trajectory, comprising all candidates during an agent's complete session for a single kernel task.
We replay each trajectory under native, \sysname{}, and ungated
on H20 and rank its candidates by their measured kernel durations.
We use the first native replay as the reference for all comparisons,
and compare a second native replay, \sysname{}, and ungated against it.
The second native replay (native repeat) captures run-to-run variation.
Figure~\ref{fig:rank} reports these comparisons at the pairwise and
trajectory levels.
For each candidate pair within a trajectory, we compute its
native kernel-duration gap from the first native replay:
the absolute difference between the two candidates' durations
divided by the smaller one.
For example, a pair with durations of 1.0 and 1.1\,ms in the first
native replay has a $10\%$ gap.
Figure~\ref{fig:rank}a groups all such pairs by this gap and reports
the fraction of pairs whose ordering differs from the reference ordering.
Grouping by gap distinguishes near-tied pairs, whose orderings are noise-sensitive,
from pairs with larger gaps.
\sysname{}'s flip rate stays within three percentage points of the native
repeat in every group.
Figure~\ref{fig:rank}b plots the CDF of per-trajectory Kendall's
$\tau_b$~\cite{kendall38-tau,kendall45-ties} between
measured and reference candidate rankings.
Values closer to 1 indicate stronger agreement.
Median $\tau_b$ is 0.857 for \sysname{}, 0.820 for the native repeat,
and 0.432 for ungated.
Together, these results show that \sysname{} achieves pairwise and
trajectory-level ranking agreement close to that of the native repeat.

\subsection{Ablation and Sensitivity Study}
\label{sec:eval-ablation}
\label{sec:eval-amd}

\begin{figure}[!tbp]
  \centering
  \includegraphics{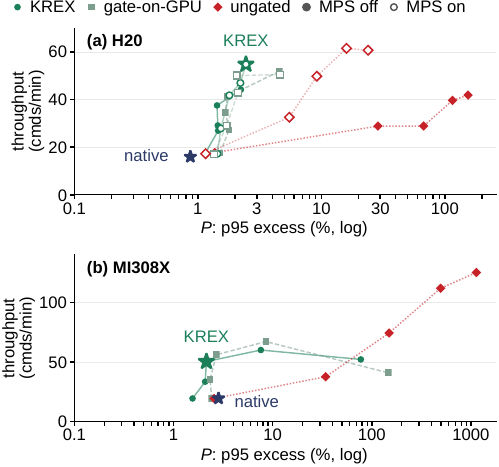}
\caption{Throughput versus $P$ (log x-axes).
Lines connect command concurrency 1, 2, 4, 8, and 16 in order; hollow markers indicate MPS on.
Stars mark native and \sysname{}.}
  \label{fig:ablation}
\end{figure}

To separate the effects of GPU protection, CPU protection, MPS, and command concurrency,
Figure~\ref{fig:ablation} compares the throughput and $P$ of ungated, gate-on-GPU,
and \sysname{} at varying command concurrency levels on H20 and MI308X.
Ungated provides no region protection; gate-on-GPU protects GPU regions only; and
\sysname{} additionally freezes sibling processes to protect host-side execution
(\S\ref{sec:design-entry}). We also compare configurations with and without MPS on H20.

On H20 at a command concurrency of 16 with MPS on, the measured $P$ is $23.87\%$
for ungated, $4.63\%$ for gate-on-GPU, and $2.45\%$ for \sysname{}, at 60.7,
50.5, and 54.8 cmds/min, respectively. GPU protection provides most of the
fidelity improvement; adding CPU protection further reduces timing inflation
while also increasing throughput in this configuration. The same fidelity
improvement is also observed on MI308X at a command concurrency of 8, with
measured $P$ of $495.3\%$, $8.6\%$, and $7.6\%$, respectively. At a command
concurrency of 2 or more, both protected configurations have lower timing
inflation than ungated on both platforms.

MPS increases \sysname{}'s throughput with little change in timing inflation. On
H20 at a command concurrency of 16, enabling MPS raises throughput from 44.2 to
54.8 cmds/min, a $24\%$ gain, while $P$ rises from $2.23\%$ to $2.45\%$.

Command concurrency determines how much throughput \sysname{} gains and how much
measurement fidelity it sacrifices. On H20 with MPS on, increasing command
concurrency from 1 to 16 raises throughput from 17.3 to 54.8 cmds/min; $P$
is $1.44$--$2.21\%$ at a command concurrency of up to 8 and reaches $2.45\%$ at
16. A command concurrency of 16 with MPS yields the highest measured \sysname{}
throughput on H20. On MI308X, command concurrency levels of 4, 8, and 16 yield
50.4, 60.1, and 52.2 cmds/min, with mean $P$ of $2.15\%$, $7.61\%$, and
$77.65\%$, respectively. A command concurrency of 8 increases throughput at
higher distortion, while 16 worsens both metrics relative to 8. We therefore
choose a command concurrency of 16 with MPS on H20 and the lower-distortion
setting of 4 on MI308X.

\subsection{Impact of GPU-Call Forwarding}
\label{sec:eval-limits}

\begin{figure}[t]
  \centering
  \includegraphics{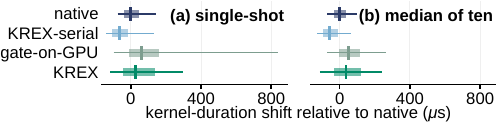}
\caption{Kernel-duration shifts relative to native's median. Ticks: medians; boxes: p25--p75; whiskers:
p05--p95. }
  \label{fig:rpc}
\end{figure}

We measure forwarding's net effect on reported kernel duration, rather than
individual RPC execution costs. Figure~\ref{fig:rpc} compares duration shifts
relative to native's median, using 896 measurements of the same candidate under
each configuration: native, \sysname{}-serial, gate-on-GPU, and \sysname{}, with
single-shot (left) and median-of-ten (right) timing. The \sysname{}-serial
median is $64\,\mu$s below native under single-shot timing and $56\,\mu$s below
under median-of-ten. These small negative shifts may reflect run-to-run variation
rather than a reduction in kernel execution time. Under single-shot timing,
gate-on-GPU and \sysname{} shift the median by $62\,\mu$s ($0.2\%$) and $28\,\mu$s
($0.1\%$), respectively. Adding CPU protection lowers the median shift under
both protocols and reduces the p95 shift from $840$ to $296\,\mu$s under
single-shot timing and from $264\,\mu$s to $238\,\mu$s under median-of-ten.

\section{Discussion and Limitations}
\label{sec:discussion}

\leadpara{Trust and region boundaries.} \sysname{} assumes cooperative candidates.
The pool's timeouts and reclamation limit abuse but do not prevent it.
Regions are marked by the benchmarking code's author, whether an agent
generating the script or a developer providing a fixed benchmarking scaffold.
\sysname{} enforces exclusivity within the marked boundaries but does not validate 
their correctness. Even with correct marks, timing fidelity is supported by empirical measurements without a theoretical worst-case error bound.

\middlepara{Device-wide control.} Some operations require control beyond 
regions. Clock settings, for example, affect all tenants and must be managed
at deployment level. When a command requires device-wide control, \sysname{}
conservatively grants it a whole-device lease, suspending pooled work on that
GPU while the command runs natively without the shim.

\middlepara{Resource scheduling and failure recovery.} \sysname{} uses slot-based
admission and provides no region-access fairness guarantee. Future work includes
scheduling by GPU memory and CPU core demands, and fair region admission to
prevent starvation when commands monopolize critical regions.
Recovery handles individual candidate failures, but dispatcher failure
interrupts service and requires restarting the pool and resubmitting in-flight
evaluations, motivating potential fail-over mechanisms.

\middlepara{Generalization.} Beyond kernel-agent benchmarking, region-granular exclusivity
could support other GPU-based self-improvement workflows that select generated
implementations using performance measurements, provided measurement boundaries
can be marked explicitly and the surrounding work tolerates sharing.
We also plan to extend \sysname{} to multi-GPU candidates and multi-node pools.
\section{Related Work}
\label{sec:related}

\PHM{GPU sharing and remoting.} GPUs are shared at three levels.
\emph{Time-sharing} alternates whole contexts on a device, at
hypervisor-chosen quanta in NVIDIA's time-sliced vGPU~\cite{nvidia-vgpu} or
with faster, finer-grained switching in software~\cite{pipeswitch-osdi20,tgs-nsdi23}.
\emph{Spatial sharing} runs tenants concurrently: MPS merges processes into
one context and MIG partitions a device into isolated
instances~\cite{nvidia-mps,nvidia-mig}, and a line of systems builds on these
to schedule, preempt, or right-size kernels for co-located
inference~\cite{reef-osdi22,paella-sosp23,orion-eurosys24,krisp-hpca23,sgdrc-ppopp25,zipbatch-socc25}.
\emph{Cluster schedulers} place and pack jobs across shared
GPUs~\cite{gandiva-osdi18,antman-osdi20,tiresias-nsdi19,gavel-osdi20,lucid-asplos23}.
\emph{API remoting} forwards GPU calls across processes, VMs, or
machines for remote access or
consolidation~\cite{rcuda-hpcs10,vcuda-ipdps09,gvirtus-europar10,cricket-jpdc22,ava-asplos20}.

\middlepara{Kernel agents.} AlphaEvolve~\cite{alphaevolve-arxiv25}, CUDA
Agent~\cite{cudaagent-arxiv26}, CUDA-L2~\cite{cudal2-arxiv25}, and
GEAK~\cite{geak-arxiv25} generate and optimize GPU kernels, while vendor work
studies how to guide kernel agents~\cite{sol-kernelagent-arxiv26}. Tensor
compilers such as TVM~\cite{tvm-osdi18} and Tensor
Comprehensions~\cite{tensor-comprehensions-arxiv18} automate kernel
optimization, and autotuners likewise search kernel spaces through
measurement~\cite{autotvm-neurips18,ansor-osdi20}. These approaches choose which
candidates to evaluate; \sysname{} addresses how to execute their benchmarking
concurrently while preserving measurement fidelity, rather than
proposing another search strategy.

\middlepara{Kernel benchmarking and measurement.} KernelBench evaluates
generated kernels on exclusive GPUs~\cite{kernelbench}, competition platforms
use dedicated runners~\cite{kernelbot-codeml25}, agent benchmarks isolate task
workspaces~\cite{agentkernelarena}, and industrial systems profile in isolated
environments~\cite{kernelevolve-arxiv25}. FastKernels~\cite{fastkernels-arxiv26}
evaluates generated kernels under production serving conditions; its focus on
compatibility with serving interfaces differs from our focus on measurement
fidelity under GPU sharing. Timing routines such as Triton's
\dobench{}~\cite{triton-mapl19} address measurement within a benchmarking
harness. Work on environmental bias~\cite{wrongdata-asplos09} and layout
randomization~\cite{stabilizer-asplos13} further highlights the importance of
measurement conditions. Statistical benchmarking methods address experimental
design and uncertainty~\cite{hoefler-sc15,kalibera-ismm13}. \sysname{}
complements these methods by excluding competing work during marked regions
while allowing concurrency elsewhere.

\section{Conclusion}
\label{sec:conclusion}

In this paper, we showed that kernel agent benchmarking need not trade GPU
utilization for measurement fidelity: only critical regions must run
alone. We presented \sysname{}, a runtime system for concurrent kernel agent
benchmarking that realizes this insight with region-granular exclusivity.
\sysname{} drains competing GPU work and freezes sibling host processes inside
regions while overlapping commands and reusing persistent GPU contexts outside
them to avoid repeated, serialized initialization. Evaluations showed that
\sysname{} delivers up to $3.4\times$ the throughput of command-granular
exclusivity at a negligible timing inflation, preserving the relative
performance signals that guide the agent's search.

\section*{Acknowledgments}
The authors used an AI assistant for prose editing. All technical content was authored and verified by the human authors.

\bibliographystyle{ACM-Reference-Format}
\bibliography{references}

\end{document}